\documentstyle[pre,aps]{revtex}

\begin{document}

\draft

\author{V.I. Yukalov, E.P. Yukalova, and V.S. Bagnato}

\address{International Centre of Condensed Matter Physics\\
University of Brasilia, CP 04513, Brasilia, DF 70919--970, Brazil \\
and\\
Instituto de Fisica de S\~ao Carlos\\
Universidade de S\~ao Paulo, CP 369, S\~ao Carlos, S\~ao Paulo 
13560--970, Brazil}
\title{Non--Ground--State Bose--Einstein Condensates of Trapped Atoms}
\maketitle

\begin{abstract}

The possibility of creating a Bose condensate of trapped atoms in a 
non--ground state is suggested. Such a nonequilibrium Bose condensate can 
be formed if one, first, obtains the conventional Bose condensate in the 
ground state and then transfers the condensed atoms to a non--ground state by 
means of a resonance pumping. The properties of ground and non--ground states 
are compared and plausible applications of such nonequilibrium condensates 
are discussed.

\end{abstract}

\vspace{1cm}


\section{Introduction}

The recent realization of Bose--Einstein condensation of dilute atomic 
gases in magnetic traps [1-3] has opened a rapidly expanding field of 
studies of condensate properties. There has been a splash of both 
experimental and theoretical activity on this subject (see books [4,5]).

Atoms trapped in a confining potential possess the discrete spectrum of 
states. At high temperatures these states are occupied, according to the 
quantum Bose--Einstein distribution, so that no state is occupied 
macroscopically. Under the macroscopic occupation of a state one implies that 
the number of atoms in this state is proportional to the total number of atoms
in the system. An important consequence of quantum statistics is that, when the
system is cooled down below some critical temperature, bosons pile up in the 
lowest energy state of a confining potential. The macroscopic population of 
the quantum--mechanical ground state of a confining potential is the 
characteristic feature of Bose--Einstein condensation.

A natural question that can be raised is: Is it possible to realize the 
macroscopic population of some other quantum--mechanical state rather 
than the ground state, or in addition to the latter? That is, can one 
produce a Bose condensate in a non--ground state? The answer to this 
question is interesting by its own. And, if that is possible, several new 
important applications can be suggested.

For example, recently two overlapping $^{87}Rb$ condensates in two 
different ground state hyperfine levels were created [6]. However, for 
other atoms the simultaneous creation of two condensates in different 
ground--state hyperfine levels may be difficult or not feasible [7]. Then 
the alternative could be the creation of two condensates, one in the 
ground--state and another in a non--ground--state level.

Realizing the macroscopic population of a non--ground state of a 
confining potential could be a way for producing various spatial 
distributions in the system of coherent atoms. This may find application 
for atom lasers for which the creation of coherent atomic beams with 
different spatial modes may be required.

One more possibility of employing such a non--ground--state condensate 
could be for studying relaxation processes in the quantum degenerate 
regime. The macroscopic occupation of an excited state would result, as 
is clear, in a nonequilibrium sample. When external forces supporting 
this state are switched off, the system will relax returning to the 
equilibrium by repopulating the discrete levels of the potential. The 
observation of this process of relaxation in the quantum degenerate 
regime can provide useful information about this form of quantum matter.

Finally, when a new system with unusual features is explored, there is 
always the chance of finding something completely unexpected.

In this paper we describe a possible way of transferring the macroscopic 
number of atoms from the conventional ground--state condensate to a 
non--ground--state level of the confining potential, thus creating a 
non--ground--state condensate.

\section{Resonance Pumping}

Assume that the Bose gas of neutral atoms has been cooled down so that 
all atoms are in a coherent condensate state. The latter is described by 
the nonlinear Schr\"odinger equation which is often called the 
Gross--Ginsburg--Pitaevskii equation [8-12]. This equation writes
\begin{equation}
i\hbar\frac{\partial\varphi}{\partial t} =\hat H\varphi ,
\end{equation}
where the nonlinear Hamiltonian
\begin{equation}
\hat H = H(\varphi) +V_p
\end{equation}
contains the nonlinear part
\begin{equation}
H(\varphi) =-\frac{\hbar^2}{2m}\nabla^2 +U_c(\stackrel{\rightarrow}{r}) +
A|\varphi|^2
\end{equation}
and, in general, a time--dependent part $V_p$ related to external fields. 
The atom--atom interaction is modelled by the $s$--wave scattering 
interaction with the amplitude
\begin{equation}
A =(N-1)4\pi\hbar^2\frac{a}{m}
\end{equation}
in which $N$ is the number of atoms in the system, $a$ is the $s$--wave 
scattering length, and $m$ is the atomic mass. The term $U_c$ is a 
confining potential. The wave function is normalized to unity: 
$(\varphi,\varphi)=1$. Assume that at the initial time $t=0$, all atoms 
are in the ground state
\begin{equation}
\varphi(\stackrel{\rightarrow}{r},0) =\varphi_0(\stackrel{\rightarrow}{r})
\end{equation}
corresponding to the minimal energy level of the eigenvalue problem
\begin{equation}
H(\varphi_n)\varphi_n = E_n\varphi_n ,
\end{equation}
in which $n$ is a multi--index enumerating quantum states. The chemical 
potential is incorporated into the notation of the energy levels $E_n$. 
Temperature is assumed to be much lower than the condensation temperature, 
since only then it is possible to condensate almost all atoms in the ground 
state. Note that the nonlinear Schr\"odinger equation (1) describes 
coherent states [13].

Since the atoms are assumed to be initially condensed in the ground 
state, to transfer them to higher levels one needs to apply an external 
pumping field that we take in the form
\begin{equation}
V_p =V(\stackrel{\rightarrow}{r})\cos\omega t .
\end{equation}
Such a field can be realized by a special modulation of the magnetic 
field producing the trap. As far as our aim is to populate a separate energy 
level, 
say a particular level $p$ with the energy $E_p$, we have to choose the 
frequency of the pumping field (7) satisfying some resonance conditions.

Denote the transition frequencies $\omega_{mn}$ by the relation
\begin{equation}
\hbar\omega_{mn} \equiv E_m -E_n
\end{equation}
and the detuning from the chosen particular transition frequency 
$\omega_{p0}$ as
\begin{equation}
\Delta\omega\equiv \omega-\omega_{p0} .
\end{equation}
The first evident resonance condition is that the detuning must be small 
as compared to the transition frequency $\omega_{p0}$ corresponding to 
the transition from the ground state, with the energy $E_0$, to the 
chosen particular state with the energy $E_p$,
\begin{equation}
\left |\frac{\Delta\omega}{\omega_{p0}}\right | \ll 1 .
\end{equation}
In addition, it is necessary that the pumping would not influence the 
neighboring states, that is the detuning must satisfy the inequalities
\begin{equation}
\left |\frac{\Delta\omega}{\omega_{p+1,p}}\right | \ll 1 , \qquad
\left |\frac{\Delta\omega}{\omega_{p,p-1}}\right | \ll 1 .
\end{equation}

The resonance conditions (10) and (11) are necessary but not yet 
sufficient, if we keep in mind to populate only one particular level. 
This goal can be reached only when the pumping does not force transitions 
between other states, which can be expressed as the inequality
\begin{equation}
\left |\frac{\Delta\omega}{\omega-\omega_{mn}}\right |\ll 1 \qquad
(m\neq p,\; n\neq 0) .
\end{equation}

The conditions (10) and (11) are easy to accomplish making the detuning 
sufficiently small. The resonance condition (12) is more restrictive 
requiring that the spectrum $\{ E_n\}$, defined by the eigenproblem (6), 
would not be equidistant, as it happens for a simple harmonic oscillator. 
Really, if that would be the case, then the pumping of atoms from the ground 
state to the chosen particular state would, at the same time, induce 
transitions from the latter to another equidistant state, and from the latter 
to higher equidistant states, and so on. In such a case, all atoms 
will be dispersed over a number of states making it impossible to get a 
macroscopic population of any of them. Fortunately, because of the 
nonlinearity, representing atomic interactions, in the Hamiltonian (3), 
the spectrum $\{ E_n\}$ is not equidistant even when the confining potential 
$U_c$ is harmonic. In addition, we may include into the confining potential 
anharmonic terms and regulate its spectral characteristics by varying 
anharmonicity parameters [14-16]. Moreover, as we shall show in Sec.III, the 
spectral properties of the nonlinear Hamiltonian (3) may be essentially 
modified by varying the intensity of interactions (4), for which it is 
sufficient to change the number of atoms $N$. Therefore, it is always possible
to prepare the system for which condition (12) holds true. The situation 
here is similar to the problem of inducing resonant electron transitions 
in an atom. The latter also contains many electron levels, but, because 
these are not equidistant, it is practically always possible to induce a 
resonant transition between a chosen pair of them [17]. The principal 
difference between the resonant electronic transitions in an atom and 
atomic transitions in a confining potential is that electronic levels are 
not equidistant because of the hydrogen--type potential, while the 
interactions between electrons do not play essential role. In such a 
case, the resonant electronic transitions can be treated in a linear 
approximation. As to the atoms in a confining potential, if the latter is 
harmonic, then the main role of making the energy levels non--equidistant 
is played by the atomic interactions. This makes the problem principally 
nonlinear and forces to deal with complicated nonlinear equations.

To describe the time evolution of the system, we have to consider the 
time--dependent nonlinear Schr\"odinger equation (1). Present its 
solution as an expansion
\begin{equation}
\varphi(\stackrel{\rightarrow}{r},t) =\sum_n c_n(t)
\varphi_n(\stackrel{\rightarrow}{r})\exp\left (-\frac{i}{\hbar}E_nt\right )
\end{equation}
in the basis of the stationary states of the eigenproblem (6). 
Substituting Eq.(13) into Eq.(1), take into account that in the double sum
$$ \sum_{k,l}c^*_kc_l(\varphi_m\varphi_k,\varphi_l\varphi_n)
\exp ( i\omega_{kl}t) $$ the main contribution comes from the term
$$ \sum_{k}|c_k|^2(\varphi_m\varphi_k,\varphi_k\varphi_n) $$
because other terms containing the oscillating factors, being summed up, 
on average cancel each other. Then from Eqs.(1) and (13) we have
\begin{equation}
i\hbar\frac{dc_n}{dt} =\sum_m\left ( V_{nm}\cos\omega t +
\sum_{k(\neq n)} A_{nkm}|c_k|^2\right ) c_me^{i\omega_{nm}t} ,
\end{equation}
with the matrix elements
$$ V_{mn} \equiv (\varphi_m,V(\stackrel{\rightarrow}{r})\varphi_n) , \qquad
A_{mkn}\equiv A(\varphi_m\varphi_k,\varphi_k\varphi_n) . $$
The solution of Eq.(14) must satisfy the normalization
$$ \sum_n|c_n(t)|^2 = 1. $$

Eq.(14) is a set of equations for the functions $c_n=c_n(t)$ enumerated 
by a multi--index $n$. Let us separate this set into the equation for the 
ground--state function $c_0$, the equation for $c_p$ representing a 
chosen state with $n=p$, and the equations for all other $c_k$ with 
$k\neq 0,p$. Introduce the notion for the population probability
\begin{equation}
n_j = n_j(t) \equiv |c_j(t)|^2 ,
\end{equation}
where the index is either $j=0$, or $j=p$, or $j=k\neq 0,p$. Also, define 
the parameters
\begin{equation}
\alpha\equiv \frac{1}{\hbar} A_{0p0} =\frac{A}{\hbar}\int
|\varphi_0(\stackrel{\rightarrow}{r})|^2
|\varphi_p(\stackrel{\rightarrow}{r})|^2d\stackrel{\rightarrow}{r}
\end{equation}
and
\begin{equation}
\beta\equiv\frac{1}{\hbar}V_{0p}=\frac{1}{\hbar}\int
\varphi_0^*(\stackrel{\rightarrow}{r})V(\stackrel{\rightarrow}{r})
\varphi_p(\stackrel{\rightarrow}{r})d\stackrel{\rightarrow}{r} .
\end{equation}

The solution to Eq.(14) can be presented as the sum of a guiding center 
plus a small oscillating ripple around the latter. The equation for the 
guiding center is obtained from Eq.(14) by averaging its right--hand side 
over time according to the rule $\frac{1}{\tau}\int_0^\tau F(t)dt$ with 
$\tau\rightarrow\infty$. During this averaging the exponential 
$e^{i\Delta\omega t}$ is treated as a constant since it is a slowly varying 
factor. Actually, at the pure resonance, when $\Delta\omega\rightarrow 0$, 
this exponential , $e^{i\Delta\omega t}\rightarrow 1$, is exactly one.

After realizing the described procedures, we obtain from Eq.(14) the system 
of equations
$$ \frac{dc_0}{dt} = -i\alpha n_pc_0 -
\frac{i}{2}\beta e^{i\Delta\omega t} c_p , $$
$$ \frac{dc_p}{dt} = -i\alpha n_0c_p -
\frac{i}{2}\beta^* e^{-i\Delta\omega t} c_0 , $$
\begin{equation}
\frac{dc_k}{dt} = 0 \qquad (k\neq 0,\; p) .
\end{equation}
Since the functions $c_j$ are complex, the system of equations in (18) 
must be completed by another system either for the complex conjugate 
functions $c_j^*$ or for the amplitudes $n_j=|c_j|^2$. The equations for 
the latter are
$$ \frac{dn_0}{dt} = Im\left (\beta e^{i\Delta\omega t}c_0^*c_p\right ) , $$
\begin{equation}
\frac{dn_p}{dt} = Im\left (\beta^* e^{-i\Delta\omega t}c_p^*c_0\right ) , 
\end{equation}
and $dn_k/dt=0$ when $k\neq 0,p$. As the initial conditions we have
\begin{equation}
c_0(0) = 1 , \qquad c_p(0) = 0 , \qquad c_k(0) = 0 .
\end{equation}

From the last of the equations in (18), together with the initial 
conditions from (20), it follows that $c_k(t)=0$ for $k\neq 0,p$. 
Therefore, the normalization condition reads
\begin{equation}
n_0(t) + n_p(t) = 1 ,
\end{equation}
which demonstrates that the atoms are concentrated in the ground state 
and $p$--level, preferentially.

Eqs.(18) and (19) form a system of complicated nonlinear differential 
equations. This system could be solved by perturbation theory in two 
limiting cases, when either  $|\alpha/\beta|\ll 1$ or $|\beta/\alpha|\ll 
1$. In the intermediate regime, when $|\alpha/\beta|\sim 1$, perturbation 
theory is not applicable. A general solution, valid for arbitrary 
relation between the parameters $\alpha$ and $\beta$, can be obtained by 
employing the method of scale separation [18-20]. This can be done by 
noticing that the functions $c_0$ and $c_p$ contain time--dependent 
imaginary factors absent in $n_0\equiv|c_0|^2$ and $n_p\equiv|c_p|^2$, 
that is, the time variation of $c_0$ and $c_p$ is faster than that of 
$n_0$ and $n_p$. Consequently, $c_0$ and $c_p$ can be classified as fast 
functions as compared to the more slow functions $n_0$ and $n_p$. Then the 
system (18) of the equations for the fast functions can be approximately
solved by keeping the slow functions, $n_0$ and $n_p$, as quasi--integrals of 
motion. From Eq.(18) we get the equations
$$ \frac{d^2c_0}{dt^2} +i(\alpha-\Delta\omega)\frac{dc_0}{dt} +
\left [\frac{|\beta|^2}{4} -\alpha n_p(\alpha n_0 -\Delta\omega)\right ] 
c_0 = 0 , $$
\begin{equation}
\frac{d^2c_p}{dt^2} +i(\alpha + \Delta\omega)\frac{dc_p}{dt} +
\left [\frac{|\beta|^2}{4} -\alpha n_0(\alpha n_p + \Delta\omega)\right ] 
c_p = 0 , 
\end{equation}
with the initial conditions (20) and
\begin{equation}
\dot c_0(0) = -i\alpha n_p , \qquad \dot c_p(0) = -\frac{i}{2}\beta^* ,
\end{equation}
where the dot means time derivative. The solution of Eqs.(22), with $n_0$ 
and $n_p$ kept fixed, writes
$$ c_0 =\left [\cos\frac{\Omega t}{2} 
+i\frac{\alpha(n_0-n_p)-\Delta\omega}{\Omega}\sin\frac{\Omega t}{2}
\right ]\exp\left\{ -\frac{i}{2}(\alpha-\Delta\omega)t\right\} , $$
\begin{equation}
c_p = -i\frac{\beta^*}{\Omega}\sin\frac{\Omega t}{2}
\exp\left\{ -\frac{i}{2}(\alpha+\Delta\omega)t\right\} ,
\end{equation}
with the effective Rabi frequency given by the expression
\begin{equation}
\Omega^2 =\left [ \alpha(n_0 -n_p) -\Delta\omega\right ]^2 +|\beta|^2 .
\end{equation}
Then for the slow functions we obtain
$$ n_0 = 1 -\frac{|\beta|^2}{\Omega^2}\sin^2\frac{\Omega t}{2} , $$
\begin{equation}
n_p = \frac{|\beta|^2}{\Omega^2}\sin^2\frac{\Omega t}{2} .
\end{equation}
The functions in Eqs.(26) describe the time evolution for the population of 
the ground state and of the chosen excited state. The form of these functions 
is similar to that one meets considering the Rabi oscillations [17]. However, 
it is worth emphasizing that, contrary to the linear case which can be 
recovered by putting $\alpha=0$, the expressions in Eqs.(26) are, actually, 
the equations for $n_0$ and $n_p$ since the effective Rabi frequency (25) 
itself depends on these populations. Because of this, the solution of the 
equations from (26) will not result in simple sinusoidal oscillations.

Consider, for example, the case when the detuning is such that it 
satisfies the relation
\begin{equation}
\alpha +\Delta\omega = 0 .
\end{equation}
Then Eq.(25) gives
\begin{equation}
\Omega =\sqrt{4\alpha^2n_0^2 +|\beta|^2} .
\end{equation}
In that case Eq.(26) shows that the ground--state level becomes empty 
while the upper resonant level completely populated, i.e.,
$$ n_0(t_k) = 0 , \qquad n_p(t_k) = 1 , $$
at the moments of time
\begin{equation}
t_k =\frac{\pi}{|\beta|}(1+ 2k) \qquad (k=0,1,2,\ldots) .
\end{equation}
As far as $n_0\rightarrow 0$, when $t\rightarrow t_k$, then the effective 
Rabi frequency (28) softens, $\Omega\rightarrow|\beta|$, and the 
motion around $t=t_k$ slows down. Hence, the system spends more time on 
the upper level than in the ground state. And if at the moment $t=t_k$ we 
switch off the pumping field (7), then we shall get an inverted system 
with all atoms being in the non--ground state. Another way of obtaining 
an inverted system could be by adiabatically varying the detuning, as in 
the regime of adiabatic passage [17], till we reach the compensation 
condition (27). The latter, certainly, has sense only when the detuning 
continues to obey the resonance conditions (10)-(12).

If the compensation condition (27) cannot be satisfied, then it is 
impossible to transfer all atoms from the ground state to the chosen 
excited state. However, it is always possible to populate these states 
equally. Really, consider the resonance case, when $\Delta\omega=0$. Then 
at the moments
\begin{equation}
t_k^* =\frac{\pi}{2|\beta|}(1 +8k) \qquad (k=0,1,2,\ldots)
\end{equation}
we have
$$ n_0(t_k^*) = n_p(t_k^*) =\frac{1}{2} , $$
that is, both states are equally populated.

It is worth paying attention to the following. The characteristic 
frequency of solutions (24) and (26) is the Rabi frequency (25). Looking 
back at Eq.(14), we see that in general, there should exist as well 
solutions with extra frequencies being combinations of basic frequencies 
entering Eq.(14). One may ask a question "when such extra frequencies 
could appear?"

Recall that solutions (24) and (26) correspond to the guiding centers 
which constitute the main approximation in the method of averaging [21] 
and in the guiding--center approach [22]. The general form of the 
population amplitudes $c_n$ is defined by Eq.(14). Denoting the general 
solution to the latter equation by $c_n^{gen}$, we may present it as a sum
$$ c_n^{gen} = c_n +\sigma_n , $$
in which $c_n$ is given by the guiding centers in (24) and $\sigma_n$ is 
an additional ripple oscillating around the guiding centers. The 
characteristic frequency of the latter is the Rabi frequency (25). The 
ripple solution $\sigma_n$ can, in turn, be written as a sum of terms 
with characteristic frequencies that are essentially higher than the Rabi 
frequency. In this way, the guiding center $c_n$ represents the main 
harmonic while the ripple solution $\sigma_n$ represents a sum of higher 
harmonics. Averaging the general solution $c_n^{gen}$ over the largest 
characteristic period corresponding to the higher harmonics gives the guiding
center $c_n$. Therefore, being interested in the average behaviour of 
solutions, one accepts the guiding center as the main approximation. 
Moreover, the ripple term $\sigma_n$ not only oscillates much faster than 
the guiding center but the amplitude of the former is smaller than that 
of the latter.

In order to concretize what is said above, let us substitute the general 
solution $c_n^{gen}=c_n+\sigma_n$ into Eq.(14). Introduce the notation
$$ \alpha_{mjn}\equiv\frac{A_{mjn}}{\hbar}, \qquad 
\beta_{mn}\equiv \frac{V_{mn}}{\hbar} . $$
Using equations in (18) for the guiding centers, we obtain the equation
$$ \frac{d\sigma_n}{dt} = - i\sum_m\left\{
\beta_{nm}\sigma_m\cos\omega t + \sum_{j(\neq n)}\alpha_{njm}\left [
|c_j|^2\sigma_m +\left ( c_j^*\sigma_m + c_j\sigma_m^*\right )
\left ( c_m +\sigma_m\right )\right ]\right\} e^{i\omega_{nm}t} $$
for the ripple term. Here $c_n$ are the guiding centers defined by 
Eqs.(18) and (24). As is evident, the equation for the ripple term 
contains various higher harmonics, as a result of which the ripple 
solution $\sigma_n$ oscillates faster than the guiding center.

Now, let us explain why the amplitude of the ripple term is smaller than that
of the guiding center. Introducing the notation 
$\varepsilon\equiv\max\{\alpha,\beta\}$, we see from (25) that the Rabi 
frequency $\Omega\sim\varepsilon$. Taking this into account and looking 
at (24), we conclude that the amplitude of the guiding center $c_n$ is of 
order unity, $c_n\sim\varepsilon/\Omega\sim 1$.

The ripple solution $\sigma_n$ can be presented as a sum of harmonics 
with amplitudes of order $\varepsilon/\Omega_\nu$, where $\Omega_\nu$ is 
a characteristic frequency of a $\nu$--harmonic. Among these 
characteristic frequencies there are various conbinations of $\omega$ and 
$\omega_{mn}\pm\omega$. Remember that, according to the quasiresonance 
condition (10), we have $\omega\sim\omega_{p0}\sim\varepsilon$. 
Consequently, the characteristic frequencies of harmonics are of 
order $\Omega_\nu\sim\nu\varepsilon$, with $\nu=2,3,\ldots$. Thus, the 
corresponding amplitudes are of order $\varepsilon/\Omega_\nu\sim 1/\nu$.
Hence, the amplitudes of higher harmonics diminish as $1/\nu$ with 
increasing $\nu=2,3,\ldots$.

In this way, we see that the guiding centers in (24) really constitute 
the main approximation to Eq.(14). In this approximation, the ripple 
solution, which oscillates faster and has smaller amplitude, can be 
neglected. If needed, the higher--order harmonics can be taken into 
account by means of perturbation theory. Such situation is common for the 
method of averaing and guiding--center approach [21,22].

The space--time distribution of atoms is given by the density
$$ \rho(\stackrel{\rightarrow}{r},t) =\rho_0(\stackrel{\rightarrow}{r},t) +
\rho_p(\stackrel{\rightarrow}{r},t) , $$
in which
$$ \rho_j(\stackrel{\rightarrow}{r},t) = Nn_j(t)
|\varphi(\stackrel{\rightarrow}{r})|^2 $$
is a partial density for $j=0,p$. These densities, are normalized to the 
total number of atoms
$$ N =\int\rho(\stackrel{\rightarrow}{r},t)d\stackrel{\rightarrow}{r} $$
and, respectively, to the number of atoms
$$ N_j =\int\rho_j(\stackrel{\rightarrow}{r},t)d\stackrel{\rightarrow}{r} 
= Nn_j(t) $$
in the corresponding states. Since the spatial dependence of the wave 
functions for different states is different, we may get condensates with 
different space distributions. In general, such condensates will coexist, 
though, if the compensation condition is achieved, a pure 
non--ground--state condensate can be realized.

In our consideration we have assumed that the system is initially cooled 
down so that all atoms are condensed in the ground state. The possible 
admixture of non--condensed thermally excited atoms has been neglected. 
Such a picture, as is known, is admissible for sufficiently low 
temperatures below the condensation point. If the temperature is kept low 
during the process of the resonant pumping, we may continue disregarding 
thermal excitations. Their role becomes important only after we switch 
off the pumping field. Since, during this pumping, the system has 
acquired additional energy, the latter can be redistributed among the 
energy levels through some relaxation mechanism. The role of such 
mechanism will be played by the interactions between condensed and 
non--condensed atoms. The thermally excited atoms will form a kind of a 
heat bath providing the possibility of relaxation to equilibrium.

\section{Stationary States}

Stationary states for the nonlinear Hamiltonian (3) are defined by the 
eigenvalue problem (6). The confining potential, typical of magnetic 
traps, is well described by the anisotropic harmonic potential
\begin{equation}
U_c(\stackrel{\rightarrow}{r}) =\frac{m}{2}(\omega_x^2 x^2 +
\omega_y^2 y^2 +\omega_z^2 z^2) .
\end{equation}

It is convenient to pass to dimensionless quantities measured in units of 
the characteristic oscillator frequency $\omega_0$ and length $l_0$ given 
by the expressions
\begin{equation}
\omega_0\equiv(\omega_x\omega_y\omega_z)^{1/3}, \qquad 
l_o\equiv\sqrt{\frac{\hbar}{m\omega_0}} .
\end{equation}
The anisotropy of potential (31) is characterized by the anisotropy 
parameters
\begin{equation}
\lambda_1\equiv\frac{\omega_x}{\omega_0} , \qquad
\lambda_2\equiv\frac{\omega_y}{\omega_0}, \qquad
\lambda_3\equiv\frac{\omega_z}{\omega_0} .
\end{equation}
Define the dimensionless coordinates
\begin{equation}
x_1\equiv\frac{x}{l_0} , \qquad x_2\equiv\frac{y}{l_0} , \qquad
x_3\equiv\frac{z}{l_0} ,
\end{equation}
forming the vector $\stackrel{\rightarrow}{x}=\{ x_1,x_2,x_3\}$. The 
dimensionless interaction parameter is
\begin{equation}
g\equiv\frac{mA}{\hbar^2l_0} = 4\pi(N-1)\frac{a}{l_0} .
\end{equation}
Introducing the dimensionless Hamiltonian and wave function, respectively,
\begin{equation}
H\equiv\frac{H(\varphi)}{\hbar\omega_0} , \qquad 
\psi(\stackrel{\rightarrow}{x})\equiv 
l_0^{3/2}\varphi(\stackrel{\rightarrow}{r}) ,
\end{equation}
we obtain
\begin{equation}
H =\frac{1}{2}\sum_{i=1}^3\left ( -\frac{\partial^2}{\partial x_i^2} 
+\lambda_i^2 x^2\right ) + g|\psi|^2 .
\end{equation}

Even when the scattering length $a$ is much less than the oscillator 
length $l_0$, the interaction parameter (35) can be very large because of 
the great number of particles $N$. This situation is similar to that 
existing for large clusters [23]. Actually, a group of atoms trapped in a 
confining potential also forms a kind of a cluster. With a large interaction
parameter, one cannot apply the standard perturbation theory in powers of 
$g$ for calculating the eigenvalues of the Hamiltonian (37). Nevertheless, 
one may employ the renormalized perturbation theory [24-30] which, as has 
been shown by a number of examples [14-16, 24-31] successfully works for 
arbitrary values of the coupling parameter, as well as for all energy 
levels, providing a good accuracy with the maximal error around $1\%$. The
first step of this approach is to choose an initial approximation containing
trial parameters that, in the following steps, will be turned into control 
functions controlling the convergence of the procedure [24]. These 
control functions are to be defined from the fixed--point and stability 
conditions [29,30]. One of the simplest forms of the fixed--point 
condition is the minimal--sensitivity condition [28] which, for the 
first--order approximation, is equivalent to the variational condition for
an energy functional [27-32].

As an initial approximation, we may take the Hamiltonian
\begin{equation}
H_0=\frac{1}{2}\sum_{i=1}^3\left (-\frac{\partial^2}{\partial x_i^2} 
+u_i^2x_i^2\right ) ,
\end{equation}
in which the effective frequencies $u_i$, with $i=1,2,3$, will be control
functions. The eigenfunctions of the Hamiltonian (38) are the oscillator 
wave functions
$$ \psi_n(\stackrel{\rightarrow}{x}) =\prod_{i=1}^3\psi_{n_i}(x_i) , \qquad
n\equiv\{ n_1,n_2,n_3\} , $$
where $n_i=0,1,2,\ldots$ Respectively, the eigenvalues of $H_0$ are given by
\begin{equation}
E_n^{(0)} =\sum_{i=1}^3u_i\left (n_i +\frac{1}{2}\right ) .
\end{equation}

Perturbation theory is to be constructed with respect to the perturbation 
$\Delta H\equiv H - H_0$, which is
$$ \Delta H =\frac{1}{2}\sum_{i=1}^3(\lambda_i^2- u_i^2)x_i^2 + g|\psi|^2 . $$ 
The eigenvalues, in the first--order approximation, are
\begin{equation}
E_n^{(1)}(\lambda,g,u) = E_n^{(0)} +\Delta E_n ,
\end{equation}
where, for compactness, the notations 
$\lambda\equiv\{\lambda_1,\lambda_2,\lambda_3\}$ and $u\equiv\{ u_1,u_2,u_3\}$
are accepted and
$$ \Delta E_n = (\psi_n,\Delta H\psi_n ) . $$

The control functions $u_i=u_i(\lambda,g,n)$ can be found from the 
variational condition
\begin{equation}
\frac{\partial}{\partial u_i} E_n^{(1)}(\lambda,g,u) = 0 ,
\end{equation}
which is a simple form of the fixed--point condition. Substituting the found
$u_i$ into Eq.(40), we have
\begin{equation}
e_n(\lambda,g) \equiv E_n^{(1)}(\lambda,g,u(\lambda,g,n)) .
\end{equation}

For what follows, it is useful to introduce the notation
\begin{equation}
g_n\equiv gJ_n, \qquad J_n\equiv\prod_{i=1}^3J_{n_i} ,
\end{equation}
in which
$$ J_{n_i} =\frac{(|\psi_{n_i}|^2,|\psi_{n_i}|^2)}{\sqrt{u_i(n)}} =
\frac{1}{\pi(2^{n_i}n_i!)^2}\int_{-\infty}^{+\infty}H_{n_i}^4(x)\exp\left (
-2x^2\right ) dx , $$
where $H_{n_i}(x)$ is a Hermite polynomial and $u_i(n)\equiv u_i(\lambda,g,n)$.

Eq.(40), with notation (43), can be written as
\begin{equation}
E_n^{(1)}(\lambda,g,u) =\frac{1}{2}\sum_{i=1}^3
\left ( n_i +\frac{1}{2}\right )\left ( u_i +\frac{\lambda_i^2}{u_i}\right ) +
\sqrt{u_1u_2u_3}\; g_n .
\end{equation}
Condition (41) results in the equation
\begin{equation}
\left ( n_i +\frac{1}{2}\right ) (u_i^2 -\lambda_i^2) +
u_i\sqrt{u_1u_2u_3} \; g_n = 0 .
\end{equation}
Using Eqs.(44) and (45), for the spectrum in Eq.(42) we have
\begin{equation}
e_n(\lambda,g) =\frac{1}{6}\sum_{i=1}^3\left ( n_i +\frac{1}{2}\right )
\left ( u_i +5\frac{\lambda_i^2}{u_i}\right ) ,
\end{equation}
where $u_i=u_i(\lambda,g,n)$ are defined by Eq.(45).

To understand better the properties of the spectrum (46), let us consider the
weak and strong coupling limits. In the weak--coupling limit, when 
$g_n\rightarrow 0$, the solution to Eq.(45) writes
\begin{equation}
u_i\simeq \lambda_i -
\frac{\sqrt{\lambda_1\lambda_2\lambda_3}}{2\left (n_i+\frac{1}{2}\right )}g_n
+\left\{ \lambda_i\frac{\sum_{j=1}^3\left ( n_j+\frac{1}{2}\right )\lambda_j
-\left ( n_i+\frac{1}{2}\right )\lambda_i}{8\prod_{j=1}^3\left ( 
n_j+\frac{1}{2}\right )} +
\frac{\lambda_1\lambda_2\lambda_3}{4\left ( n_i+\frac{1}{2}\right )^2\lambda_i}
\right\} g_n^2 .
\end{equation} 
Substituting this into Eq.(46), we get
\begin{equation}
e_n(\lambda,g) \simeq \sum_{i=1}^3\left ( n_i +\frac{1}{2}\right )\lambda_i +
\sqrt{\lambda_1\lambda_2\lambda_3}\; g_n -\frac{1}{8}\sum_{i=1}^3
\frac{\lambda_1\lambda_2\lambda_3}{\left ( n_i +\frac{1}{2}\right )\lambda_i} 
g_n^2 ,
\end{equation}
as $g_n\rightarrow 0$.

In the strong--coupling limit, when $g_n\rightarrow\infty$, the solution 
to Eq.(45) reads
\begin{equation}
u_i\simeq\frac{\left ( n_i+\frac{1}{2}\right )\lambda_i^2}
{\left [\prod_{j=1}^3\left (n_j+\frac{1}{2}\right )\lambda_j^2\right ]^{1/5}}
\; g_n^{-2/5} .
\end{equation}
And for the spectrum (46) we obtain
\begin{equation}
e_n(\lambda,g)\simeq\frac{5}{2}\left [\prod_{j=1}^3\left ( n_j +\frac{1}{2}
\right )\lambda_j^2\right ]^{1/5} g_n^{2/5} ,
\end{equation}
as $g_n\rightarrow\infty$.

\section{Ground State}

The ground state plays a special role for the phenomenon of Bose 
condensation. Therefore, we pay a little more of attention to the case of 
$n_i=0$. Since $J_0=(2\pi)^{-1/2}$, the effective interaction strength 
(43) becomes
\begin{equation}
g_0 =\frac{g}{(2\pi)^{3/2}} .
\end{equation}
Eq. (46), defining control functions, simplifies to
\begin{equation}
u_i^2 +2g_0u_i\sqrt{u_1u_2u_3}-\lambda_i^2 = 0 .
\end{equation}
And the spectrum (46) reduces to
\begin{equation}
e_0(\lambda,g) =\frac{1}{12}\sum_{i=1}^3\left ( u_i +5\frac{\lambda_i^2}{u_i}
\right ) .
\end{equation}

In the weak coupling limit, when $g_0\rightarrow 0$, Eq.(52) yields
\begin{equation}
u_i\simeq \lambda_i -\sqrt{\lambda_1\lambda_2\lambda_3}\; g_0 +\left [
\frac{\lambda_i}{2}\left ( \sum_{j=1}^3\lambda_j -\lambda_i\right ) +
\frac{\lambda_1\lambda_2\lambda_3}{\lambda_i}\right ] g_0^2 .
\end{equation}
And for the spectrum (53), we have
\begin{equation}
e_0(\lambda,g) \simeq\frac{1}{2}(\lambda_1+\lambda_2+\lambda_3) +
\sqrt{\lambda_1\lambda_2\lambda_3}\; g_0 - 
\frac{1}{16}(\lambda_1\lambda_2 +\lambda_2\lambda_3 +\lambda_3\lambda_1)g_0^2 ,
\end{equation}
as $g_0\rightarrow 0$.

In the strong--coupling limit, when $g_0\rightarrow\infty$, for the 
control functions we get
\begin{equation}
u_i\simeq 
\frac{\lambda_i^2}{(2\lambda_1\lambda_2\lambda_3)^{2/5}}g_0^{-2/5} ,
\end{equation}
and, respectively, for the spectrum we find
\begin{equation}
e_0(\lambda,g) 
\simeq\frac{5}{4}(2\lambda_1\lambda_2\lambda_3)^{2/5}g_0^{2/5} .
\end{equation}
For an axially symmetric potential, with $\lambda_1=\lambda_2\neq\lambda_3$, 
the strong  coupling limit (57) reduces to that found by Baym and Pethick 
[33].

For an isotropic potential, for which $\lambda_i=1$, there is only one 
control function given by the equation
\begin{equation}
u^2 +2g_0u^{5/2} -1 = 0 . 
\end{equation}
Then the ground state energy (53) becomes
\begin{equation}
e_0(g) =\frac{1}{4}\left ( u +\frac{5}{u}\right ) .
\end{equation}

In the weak--coupling limit, when $g_0\rightarrow 0$, Eq.(58) gives
\begin{equation}
u\simeq 1 -g_0 + 2g_0^2 .
\end{equation}
And the spectrum (59) is
\begin{equation}
e_0(g)\simeq \frac{3}{2} +g_0 -\frac{3}{4}g_0^2 ,
\end{equation}
as $g_0\rightarrow 0$.

In the strong--coupling limit, as $g_0\rightarrow\infty$, we have
\begin{equation}
u\simeq (2g_0)^{-2/5} ,
\end{equation}
and, respectively,
\begin{equation}
e_0(g)\simeq\frac{5}{4}(2g_0)^{2/5} .
\end{equation}

For the atoms with negative scattering lengths, as in the case of $^7Li$ 
or $^{85}Rb$, the interaction parameter (35) is negative. All 
weak--coupling expansions, such as Eqs.(48), (55) and (61), are the same 
for the negative $g\rightarrow -0$. When $g$ increases, the real solution 
for the spectrum exists till some critical value $g_c$ after which it 
becomes complex. Thus, for the isotropic potential, the stable ground state
is defined for $g_0^c < g_0 < 0$, with
\begin{equation}
g_0^c = -\frac{2}{5^{5/4}} = -0.267496.
\end{equation}
From here, the critical interaction parameter is
\begin{equation}
g_c = (2\pi)^{3/2}g_0^c = - 4.212960 .
\end{equation}
This, according to notation (35), defines the critical number of atoms
\begin{equation}
N_c = 1 +\frac{g_cl_0}{4\pi a}
\end{equation}
that can condense in a stable state. After $g < g_c$, the spectrum 
becomes complex. The latter means that the corresponding states are not 
any more stationary but quasistationary. The lifetime of a 
quasistationary state is defined [34,35] as
\begin{equation}
\tau_n(g) =\frac{1}{2\omega_0|Im\; e_n(g)|} .
\end{equation}

The complex spectrum of quasistationary states is defined, as earlier, by 
Eq. (46), with $u_i$ from Eq.(45). These equations, in the case of 
negative $g<0$, have several complex solutions, of which we must choose 
that continuously branching, at $g=g_c$, from a solution that is real at 
$g\rightarrow -0$. Solving these equations for the isotropic ground 
state, we find
$$ Re\; u \simeq 0.705470g^{-2/5} + 3.850396g^{-6/5} + 12.402511 g^{-2} , $$
\begin{equation}
Im \; u \simeq 2.171212 g^{-2/5} + 2.797476 g^{-6/5}
\end{equation}
for $g\gg 1$. For the real and imaginary parts of the energy we obtain
$$ Re\; e_0(g) \simeq 0.169198g^{2/5} + 0.529102 g^{-2/5} + 1.443899 
g^{-6/5} +31.006277 g^{-2} , $$
\begin{equation}
Im\; e_0(g)\simeq -0.520739g^{2/5} + 1.628409 g^{-2/5} + 1.049054 g^{-6/5} .
\end{equation}
Therefore, for the lifetime (67) we have
\begin{equation}
\tau_0(g)\simeq\frac{0.960174}{\omega_0g^{2/5}} .
\end{equation}
If the number of atoms, with a negative scattering length, exceeds the 
critical number given by Eq.(66), then only $N_c$ atoms can form a stable 
Bose condensate, excessive atoms being expelled out of the condensate 
during the time (70).

\section{Excited States}

The energy of any excited state is defined by Eq.(46). In order to 
illustrate the relative to each other behaviour of these states, let us 
write explicitly the corresponding formulas for several first levels.
Consider, for the isotropic case, the energy levels $e_{100}(g), \; 
e_{110}(g)$ and $e_{200}(g)$. For these levels, the integrals $J_{n_i}$ 
entering notation (43) are
$$ J_0 =\frac{1}{\sqrt{2\pi}}, \qquad J_1 =\frac{3}{4\sqrt{2\pi}}, \qquad
J_2 =\frac{41}{64\sqrt{2\pi}} . $$

In the weak--coupling limit, the energies of the first three excited 
states, compared to that of the ground state, behave as
$$ e_0(g)\simeq \frac{3}{2} + 0.063494g , $$
$$ e_{100}(g)\simeq \frac{5}{2} + 0.047620g , $$
$$ e_{110}(g)\simeq \frac{7}{2} + 0.035715g , $$
\begin{equation}
e_{200}(g)\simeq \frac{7}{2} + 0.040676g , 
\end{equation}
when $g\rightarrow 0$. And in the strong--coupling limit, we get
$$ e_0(g)\simeq 0.547538 g^{2/5} , $$
$$ e_{100(}g)\simeq 0.607943g^{2/5} , $$
$$ e_{200}(g)\simeq 0.632193 g^{2/5} , $$
\begin{equation}
e_{110}(g)\simeq 0.675012 g^{2/5} , 
\end{equation}
as $g\rightarrow\infty$.

Notice an interesting and important fact that the energy levels 
$e_{110}(g)$ and $e_{200}(g)$ cross each other as $g$ varies. Really, 
from Eq.(71) it follows that $e_{110}(g) <e_{200}(g)$ at small $g$, while 
Eq.(72) shows that $e_{110}(g) > e_{200}(g)$ at large $g$. This crossing 
of levels demonstrates that the latter cannot be classified being based on 
the harmonic oscillator spectrum. To correctly classify the levels, one has
to calculate their energies for each given interaction strength $g$. At 
the same time, the strong distortion of the harmonic--oscillator spectrum 
plays a positive role for our purpose, making the spectrum of interacting 
atoms nonequidistant, which is necessary for the possibility of 
transferring the ground--state Bose condensate to a non--ground state, as 
is explained in Sec.II.

The transition frequency, measured in units of $\omega_0$, is given by 
the difference
\begin{equation}
\omega_n(g) \equiv e_n(g) - e_0(g) .
\end{equation}
For the low--lying excited states this leads to
$$ \omega_{100} \simeq 1 -0.015874 g , $$
$$ \omega_{110} \simeq 2 -0.027779 g , $$
\begin{equation}
\omega_{200} \simeq 2 -0.022818 g , 
\end{equation}
in the weak--coupling limit, $g\rightarrow 0$, and to
$$ \omega_{100}(g) \simeq 0.060405 g^{2/5} , $$
$$ \omega_{200}(g) \simeq 0.084655 g^{2/5} , $$
\begin{equation}
\omega_{110}(g) \simeq 0.127474 g^{2/5} , 
\end{equation}
in the strong--coupling limit, as $g\rightarrow\infty$. Eqs.(71)-(75) show 
that sufficiently strong interaction parameter $g$ makes the spectrum 
nonequidistant, which favours the possibility of the resonance pumping of 
atoms from the ground state to an excited level.

The shape of the atomic cloud, in a given state, can be characterized by 
the aspect
ratio
\begin{equation}
R_{13}\equiv
\left (\frac{\langle x_1^2\rangle}{\langle x_3^2\rangle}\right )^{1/2} ,
\end{equation}
in which $\langle x_1^2\rangle$ is a mean--square deviation in the 
$x$--direction, and $\langle x_3^2\rangle$ is a mean--square deviation in 
the $z$--direction. Similarly, we may define the aspect ratio $R_{23}$. 
In the case of cylindrical symmetry, $R_{13}=R_{23}$.

The aspect ratio (76) can be written as
\begin{equation}
R_{13} =\left [\frac{\left ( n_1 +\frac{1}{2}\right ) u_3}{\left ( n_3
+\frac{1}{2}\right ) u_1}\right ]^{1/2} .
\end{equation}
In the weak--coupling limit this yields
\begin{equation}
R_{13}\simeq \sqrt{
\frac{\left ( n_1+\frac{1}{2}\right )\lambda_3}
     {\left ( n_3+\frac{1}{2}\right )\lambda_1}}
\left\{ 1 +\frac{\sqrt{\lambda_1\lambda_2\lambda_3}}{4(2\pi)^{3/2}}\left [
\frac{1}{\left ( n_1 +\frac{1}{2}\right )\lambda_1} -
\frac{1}{\left ( n_3 +\frac{1}{2}\right )\lambda_3}\right ] g\right \} ,
\end{equation}
when $g\rightarrow 0$. For example, for the ground state, Eq.(78) reduces to
\begin{equation}
R_{13}^0\simeq\sqrt{\frac{\lambda_3}{\lambda_1}}\left\{ 1 +
\frac{\sqrt{\lambda_1\lambda_2\lambda_3}}{2(2\pi)^{3/2}}\left (
\frac{1}{\lambda_1} - \frac{1}{\lambda_3}\right ) g\right \} .
\end{equation}
In the strong--coupling limit, Eq.(77) gives
\begin{equation}
R_{13} \simeq \frac{\lambda_3}{\lambda_1} \qquad (g\rightarrow \infty)
\end{equation}
for all energy levels.

For the ground state, our results agree with the aspect ratio found by 
other authors [33,36,37]. And for excited states, we predict an 
interesting fact that the aspect ratio in the strong--coupling limit 
becomes asymptotically, as $g\rightarrow\infty$, independent of level 
numbers, according to eq.(80).

\section{Interaction Amplitude}

An important quantity defining whether it is possible to transfer all 
atoms from the ground state to an upper level, or only half of them, is 
the interaction amplitude (16). For convenience, let us define the 
dimensionless amplitude
\begin{equation}
\alpha_n(g)\equiv\frac{\alpha}{\omega_0} ,
\end{equation}
which may be presented as
$$ \alpha_n(g) =\sqrt{u_1(0)u_2(0)u_3(0)}\; I_n g , $$
where $u_i(n)\equiv u_i(\lambda,g,n)$ and
$$ I_n\equiv\prod_{i=1}^3I_{n_i} , \qquad n\equiv\{ n_1,n_2,n_3\} , $$
$$ I_{n_i}\equiv\frac{(|\psi_0|^2,|\psi_{n_i}|^2)}{\sqrt{u_i(0)}} =
\frac{1}{\pi2^{n_i}n_i!}\int_{-\infty}^{+\infty} H_{n_i}^2(x)\exp\left\{
-\frac{u_i(0)+u_i(n)}{u_i(n)}x^2\right\} dx . $$
The last integral can be expressed as
$$ I_{n_i} =\frac{(-1)^{n_i}\Gamma\left ( n_i+\frac{1}{2}\right )}
{\pi n_i!\zeta_i^{n_i}}\sqrt{\frac{2\zeta_i-1}{2\zeta_i}}
F\left ( -n_i,-n_i;\frac{1}{2}-n_i;\zeta_i\right ) $$
through the gamma function $\Gamma(\ldots)$ and the hypergeometric 
function $F(\ldots)$, with
$$ \zeta_i\equiv\frac{u_i(0)+u_i(n)}{2u_i(0)} . $$

In the weak--coupling limit, using the equality
$$ F\left ( -n_i,-n_i;\frac{1}{2}-n_i;1\right ) = (-1)^{n_i} , $$
we find
\begin{equation}
\alpha_n(g) \simeq\frac{\sqrt{2\lambda_1\lambda_2\lambda_3}}{4\pi^3}
\left [
\prod_{i=1}^3\frac{\Gamma\left ( n_i+\frac{1}{2}\right )}{n_i!}\right ] g ,
\end{equation}
as $g\rightarrow 0$. And in the strong--coupling limit, we obtain
$$ \alpha_n(g)\simeq (\lambda_1\lambda_2\lambda_3)^{2/5}
\left (\frac{64}{\pi^7}\right )^{3/10}\times $$
\begin{equation}
\times \left [ 
\prod_{i=1}^3\frac{(-1)^{n_i}}{n_i!}\Gamma\left (n_i +\frac{1}{2}\right )
\sqrt{\frac{2\zeta_i-1}{2\zeta_i}}
F\left ( -n_i,-n_i;\frac{1}{2}-n_i;\zeta_i\right ) \right ]g^{2/5} ,
\end{equation}
where $g\rightarrow\infty$ and
$$ \zeta_i =\frac{1}{2} +\frac{n_i+\frac{1}{2}}{2\pi\left [ 8\prod_{i=1}^3
\left ( n_i +\frac{1}{2}\right ) J_{n_i}^2\right ]^{1/5}} . $$

To compare the interaction amplitudes for the first several levels with 
the corresponding energies and transition frequencies studied in Sec.V, 
let us take the isotropic case. Then, in the weak coupling limit, Eq.(82) 
yields
$$ \alpha_{000}(g)\simeq 0.06349 g , $$
$$ \alpha_{100}(g)\simeq 0.03175 g , $$
$$ \alpha_{110}(g)\simeq 0.01587 g , $$
\begin{equation}
\alpha_{200}(g)\simeq 0.02381 g ,
\end{equation}
as $g\rightarrow 0$. And in the strong--coupling limit, from Eq.(83) we 
derive
$$ \alpha_{000}(g)\simeq 0.21902 g^{2/5} , $$
$$ \alpha_{100}(g)\simeq 0.18304 g^{2/5} , $$
$$ \alpha_{110}(g)\simeq 0.14759 g^{2/5} , $$
\begin{equation}
\alpha_{200}(g)\simeq 0.17559 g^{2/5} , 
\end{equation}
as $g\rightarrow\infty$.

Comparing the interaction amplitudes (84) and (85) with the transition 
frequencies (74) and (75), we see that the compensation condition (27) 
could be accomplished in the weak--coupling limit or in the intermediate 
region of $g$, where $\alpha_n(g)\ll\omega_n(g)$. In the strong--coupling 
limit, $\alpha_n(g)$ becomes of the order of $\omega_n(g)$ and condition 
(27) cannot be satisfied, since $|\Delta\omega|\ll\omega_n(g)$. This means 
that, with increasing $g$, that is, with increasing the number of atoms 
in the ground--state condensate, it becomes more difficult to transfer 
all these atoms to an upper level. For large atomic clouds, only a half 
of the condensed atoms can be pumped up to an upper level.

\section{Discussion}

We suggested a mechanism for creating nonequilibrium Bose condensates in 
non--ground states. The possibility of transferring either all atoms or 
only a part of them to a non--ground state depends on the parameters of 
the system. In principle, these parameters are changeable, and we think 
that it is possible to adjust them so that to effectively realize the 
mechanism suggested. For concreteness, let us look at the parameters 
typical of magnetic traps used now for condensing atoms in the ground state.

In the JILA trap [1,38] the atoms of $^{87}Rb$, with the mass 
$m=1.445\times 10^{-22}g$ and the scattering length $a=6\times 
10^{-7}cm$, were condensed. The characteristic frequencies of the 
confining potential are $\omega_x=\omega_y=2\pi\times 132\; Hz$ and 
$\omega_z=2\pi\times 373\; Hz$, their geometric mean, defined in Eq.(32), 
being $\omega_0=2\pi\times 187\; Hz$. The anisotropy parameters (33) are 
$\lambda_1=\lambda_2=\frac{1}{\sqrt{2}}$ and $\lambda_3=2$. The oscillator
length $l_0=0.788\times 10^{-4}cm$. Therefore, the interaction parameter 
(35) is
$$ g= 9.565(N-1)\times 10^{-2} . $$
For the number of atoms $N\sim 10^3$, one has $g\sim 100$, which 
corresponds to the strong--coupling limit.

In the MIT trap [3,39] a condensate of $^{23}Na$, with $m=0.382\times 
10^{-22}g$ and $a=5\times 10^{-7}cm$, was realized. The characteristic 
frequencies are $\omega_x=\omega_y=2\pi\times 320\; Hz$ and 
$\omega_z=2\pi\times 18\; Hz$, so that their geometric mean is 
$\omega_0=2\pi\times 123\; Hz$. The anisotropy parameters are 
$\lambda_1=\lambda_2=2.610$ and $\lambda_3=0.147$. The oscillator length 
is $l_0=1.890\times 10^{-4}cm$. Thus, the interaction parameter (35) is
$$ g=3.324(N-1)\times 10^{-2} . $$
For $N\sim 5\times 10^5-5\times 10^6$, this gives $g\sim 10^4-10^5$, 
which certainly is in the strong coupling limit.

In the RQI trap [2,40] the atoms of $^7Li$, having $m=0.116\times 
10^{-22}g$ and the negative scattering length $a=-1.5\times 10^{-7}cm$,
were cooled down to the Bose condensation regime. With the characteristic 
frequencies $\omega_x=2\pi\times 150.6\; Hz,\;\omega_y=2\pi\times 152.6\; Hz$,
and $\omega_z=2\pi\times 131.5\; Hz$, the geometric--mean frequency is 
$\omega_0=2\pi\times 145\; Hz$. The anisotropy parameters (33) are 
$\lambda_1=1.039,\;\lambda_2=1.052$, and $\lambda_3=0.907$. The 
oscillator length $l_0=3.160\times 10^{-4}cm$. The interaction parameter
(35) becomes
$$ g=-0.597(N-1)\times 10^{-2} . $$
Since the confining potential is almost isotropic 
$(\lambda_1\approx\lambda_2\approx\lambda_3)$ we may take the value (65) 
for the critical interaction strength allowing yet a stable condensate. 
Then, the critical number (66) of atoms allowed in the condensate is 
$N_c=707$, which is in agreement with the estimates of other authors 
[33,37] giving $N_c\sim10^3$.

In this way, the trapped clouds of $^{87}Rb$ and $^{23}Na$ corresponds to 
the strong--coupling limit, and that of $^7Li$, to an intermediate 
regime, with $g\leq 4$. Therefore, if the number of condensed atoms in a 
cloud is $N\geq 10^3$, it would be difficult to transfer all of them to an 
upper level. However, as is discussed in Sec.II, it is always possible to 
transfer a half of the ground--state atoms to a non--ground--state level, 
thus, creating two coexisting condensates of the same species in two 
different quantum states. In addition, it is, probably, possible to construct
specially designed traps allowing to pump up all atoms from the ground 
state to another level, forming a pure non--ground--state Bose condensate.
In any case, both these possibilities, being realized, would produce a new 
kind of quantum ultra--cold matter which may reveal interesting and 
unexpected properties.

\vspace{7mm}

{\bf Acknowledgement}

\vspace{2mm}

We are grateful to B.W. Shore for very useful discussions. Financial 
support from the National Science and Technology Development Council of 
Brazil and from the S\~ao Paulo State Research Foundation is 
appreciated. One of us (VSB) also acknowledges support from the program 
PRONEX.


\begin{references}
\bibitem{1}
M.H. Anderson, J.R. Ensher, M.R. Matthews, C.E. Wieman, and E.A. Cornell,
Science {\bf 269}, 198 (1995).
\bibitem{2}
C.C. Bradley, C.A. Sackett, J.J. Tollett, and R.G. Hulet, Phys. Rev. 
Lett. {\bf 75}, 1687 (1995).
\bibitem{3}
K.B. Davis, M.O. Mewes, M.R. Andrews, N.J. van Druten, D.S. Durfee, D.M. 
Kurn, and W. Ketterle, Phys. Rev. Lett. {\bf 75}, 3969 (1995).
\bibitem{4}
{\it Book of Abstracts of Workshop on Collective Effects in Ultracold 
Atomic Gases}, Les Houches, France (1996).
\bibitem{5}
{\it Bose--Enstein Condensation}, Special Issue of J. Res. Natl. Inst. 
Stand. Technol. {\bf 101}, N4 (1996).
\bibitem{6}
C.J. Myatt, E.A. Burt, R.W. Christ, E.A. Cornell, and C.E. Wieman, Phys. 
Rev. Lett. {\bf 78}, 586 (1997).
\bibitem{7}
S.J. Kokkelmans, H.M. Boesten, and B.J. Verhaar, Phys. Rev. A {\bf 55}, 
1589 (1997).
\bibitem{8}
E.P. Gross, Phys. Rev. {\bf 106}, 161 (1957).
\bibitem{9}
V.L. Ginzburg and L.P. Pitaevskii, J. Exp. Theor. Phys. {\bf 7}, 858 (1958).
\bibitem{10}
E.P. Gross, Nuovo Cimento {\bf 20}, 454 (1961).
\bibitem{11}
L.P. Pitaevskii, J. Exp. Theor. Phys. {\bf 13}, 451 (1961).
\bibitem{12}
E.P. Gross, J. Math. Phys. {\bf 4}, 195 (1963).
\bibitem{13}
A.C. Biswas and S.R. Shenoy, Physica B {\bf 90}, 265 (1977).
\bibitem{14}
E.P. Yukalova and V.I. Yukalov, Phys. Lett. A {\bf 175}, 27 (1993).
\bibitem{15}
E.P. Yukalova and V.I. Yukalov. J. Phys. A {\bf 26}, 2011 (1993).
\bibitem{16}
V.I. Yukalov and E.P. Yukalova, Laser Phys. {\bf 5}, 154 (1995).
\bibitem{17}
B.W. Shore, {\it The Theory of Coherent Atomic Excitation} (John Wiley, 
New York, 1990).
\bibitem{18}
V.I. Yukalov, Phys. Rev. Lett. {\bf 75}, 3000 (1995).
\bibitem{19}
V.I. Yukalov, Laser Phys. {\bf 5}, 970 (1995).
\bibitem{20}
V.I. Yukalov, Phys. Rev. B {\bf 53}, 9232 (1996).
\bibitem{21}
N.N. Bogolubov and Y.A. Mitropolsky, {\it Asymptotic Methods in the 
Theory of Nonlinear Oscillations} (Gordon and Breach, New York, 1961).
\bibitem{22}
N.G. Van Kampen, Phys. Rep. {\bf 124}, 69 (1985).
\bibitem{23}
V.I. Yukalov and E.P. Yukalova, Physica A {\bf 223}, 15 (1996).
\bibitem{24}
V.I. Yukalov, Mosc. Univ. Phys. Bull. {\bf 31}, 10 (1976).
\bibitem{25}
W.E. Caswell, Ann. Phys. {\bf 123}, 152 (1979).
\bibitem{26}
I. Halliday and P. Surranyi, Phys. Rev. D {\bf 21}, 1529 (1980).
\bibitem{27}
J. Killingbeck, J. Phys. A {\bf 14}, 1005 (1981).
\bibitem{28}
P.M. Stivenson, Phys. Rev. D {\bf 23}, 2916 (1981).
\bibitem{29}
V.I. Yukalov, J. Math. Phys. {\bf 32}, 1235 (1991).
\bibitem{30}
V.I. Yukalov, J. Math. Phys. {\bf 33}, 3994 (1992).
\bibitem{31}
I.D. Feranchuk, L.I. Komarov, I.V. Nichipor, and A.P. Ulyanenkov, Ann. 
Phys. {\bf 238}, 370 (1995).
\bibitem{32}
R.J. Dodd, J. Res. Natl. Inst. Stand. Technol. {\bf 101}, 545 (1996).
\bibitem{33}
G.Baym and C.J. Pethik, Phys. Rev. Lett. {\bf 76}, 6 (1996).
\bibitem{34}
R. Newton, {\it Scattering Theory of Waves and Particles} (McGraw--Hill, 
New York, 1966).
\bibitem{35}
V.I. Melnikov, Phys. Rep. {\bf 209}, 1 (1991).
\bibitem{36}
M. Edwards, R.J. Dodd, C.W. Clark, P.A. Ruprecht, and K. Burnett, Phys. 
Rev. A {\bf 53}, 1950 (1996).
\bibitem{37}
F. Dalfovo and S. Stringari, Phys. Rev. A {\bf 53}, 2477 (1996).
\bibitem{38}
J.R. Ensher, D.S. Jin, M.R. Matthews, C.E. Wieman, and E.A. Cornell, 
Phys. Rev. Lett. {\bf 77}, 4984 (1996).
\bibitem{39}
M.O. Mewes, M.R. Andrews, N.J. van Druten, D.M. Kurn, D.S. Durfee, and
W. Ketterle, Phys. Rev. Lett. {\bf 77}, 416 (1996).
\bibitem{40}
C.C. Bradley, C.A. Sackett, and R.G. Hulet, Phys. Rev. Lett. {\bf 78}, 
985 (1997).
\end{references}
\end{document}